\documentstyle[preprint,aps]{revtex}

\begin{document}
\draft

\title{
	Spectroscopic factors for bound s--wave states
	derived from neutron scattering lengths
}

\author{
        P.~Mohr, H.~Herndl, and H.~Oberhummer
}

\address{
        Institut f\"ur Kernphysik, Technische Universit\"at Wien,
        Wiedner Hauptstra{\ss}e 8--10, A--1040 Wien, Austria
}

\date{\today}

\maketitle

\begin{abstract}
A simple and model--independent method is described to derive
neutron single--particle spectroscopic factors of bound $s$--wave
states in $^{A+1}Z$ = $^{A}Z \otimes {\rm{n}}$ nuclei 
from neutron
scattering lengths. Spectroscopic factors for the nuclei 
$^{13}$C, $^{14}$C, $^{16}$N,
$^{17}$O, $^{19}$O, $^{23}$Ne, $^{37}$Ar, and $^{41}$Ar
are compared to results derived from transfer experiments 
using the well--known DWBA analysis and to shell model calculations.
The scattering length of $^{14}$C is calculated from the
$^{15}$C$_{\rm{g.s.}}$ spectroscopic factor.
\end{abstract}


\narrowtext

Spectroscopic factors (SF) are an important ingredient for the
calculation of direct transfer reaction cross sections in the 
Distorted Wave Born Approximation (DWBA)
and capture reaction cross sections in the Direct Capture (DC) model.
Usually, SF can be determined experimentally by the ratio of the measured
transfer reaction cross section to the cross section calculated with DWBA
\begin{equation}
C^2 S_i = \sigma_i^{\rm{exp}} \, / \, \sigma_i^{\rm{DWBA}}
\label{sf:dwba}
\end{equation}
for each final state $i$. In the case of neutron transfer mainly
(d,p) reactions were analyzed to determine the neutron single--particle SF.
This determination has relatively large uncertainties because the optical
potentials of both the entrance and the exit channel have to be known
accurately for a reliable DWBA calculation. Usually one obtains SF with
uncertainties of up to 20\%. However, 
in many cases systematic deviations
exceeding the claimed uncertainties can be found when the results
of various experiments
(different transfer reactions like (d,p), ($^4$He,$^3$He), ($^7$Li,$^6$Li),
etc.~at different energies) 
are compared (see, e.g., Table 8 of Ref.~\cite{cook87} or
Table II of Ref.~\cite{yasue92}).

Recently, our group showed that a model--independent method exists
to extract SF from the thermal neutron capture cross section
\cite{beer_ca48}:
\begin{equation}
C^2 S_i = \sigma_{i}^{\rm{exp}}({\rm{n}}_{\rm{th}},\gamma) \, / \,
	\sigma_{i}^{\rm{DC}}({\rm{n}}_{\rm{th}},\gamma)
\label{eq:sfdc}
\end{equation}
This method has very limited uncertainties because 
at thermal energies the neutron optical
potential can be adjusted properly to the scattering length.
However, because the thermal (n,$\gamma$) 
cross section is dominated 
by incoming s--waves and E1 transitions, 
this procedure works well only for bound p--waves in the
residual nucleus.

In this work we present a simple and model--independent procedure
for the extraction of SF of bound s--waves from the free scattering
length $b$.
In this work we use the free nuclear scattering length $b$, which is
related to the bound scattering length by 
$b = ( b_{\rm{bound}} - Z \cdot b_{ne} ) \cdot A / (A+1)$
with the neutron--electron interaction length 
$b_{ne} = (-1.38 \pm 0.03) \cdot 10^{-3}~{\rm{fm}}$
\cite{mu81,koester76}.
These SF are very important for the calculation of the
(n,$\gamma$) cross section at astrophysically relevant energies 
in the order of several keV
where transitions from incoming p--waves to bound s-- and d--waves
become comparable to the transitions from the incoming s--wave to bound
p--waves~\cite{bee96,iga95}.

The method can be applied to light and intermediate nuclei with only one bound
s--wave or a strong s--wave state 
close to the neutron separation threshold.
In these cases the scattering length can be interpreted as the
very broad positive--energy wing of the s--wave subthreshold state.
The comparison of the calculated width assuming a single particle
configuration and the experimental width of this
subthreshold state leads to the SF:
\begin{equation}
C^2 S = \Gamma^{\rm{exp}} \, / \,
	\Gamma^{\rm{calc}}_{\rm sp}
\label{eq:sfwidth}
\end{equation}
This calculation is performed in the following way:

First, the wave function of the
subthreshold state is calculated using a neutron--nucleus optical potential.
The potential strength (parameters $V_0$ or $\lambda$, see below)
is adjusted to reproduce the binding energy
of the bound state (taking into account the Pauli principle by
$q = 2n + l$ where $q$, $n$, and $l$ are the oscillator, radial node, and 
angular momentum quantum numbers). In this work both Woods--Saxon (WS)
\begin{equation}
V_{\rm WS}(r) = V_0 \cdot ( 1 + \exp{(r-R/a)} )^{-1}, \; \; \;
\label{eq:WS}
\end{equation}
with $R = R_0 \cdot A_{\rm T}^{1/3}$, 
$R_0 = 1.25~{\rm{fm}}$, and
$a = 0.65~{\rm{fm}}$,
and folding potentials \cite{devries87,kobos84,abele93}
\begin{equation}
V_{\rm F}(r) = 
        \lambda
	\int \int \rho_{\rm P}(r_{\rm P}) \, \rho_{\rm T}(r_{\rm T}) \, 
v_{\rm eff}(s,\rho,E) \; 
		d^3r_{\rm P} \; d^3r_{\rm T}
\label{eq:DF}
\end{equation}
were used; the results practically do not depend on the chosen
parameterization of the optical potential. In this sense this method is
model--independent.

Second, we calculate the single particle scattering length 
$b^{\rm calc}_{\rm sp}$ 
and the width $\Gamma^{\rm calc}_{\rm sp}$ from the
optical potential which was adjusted to the bound state energy $E_B$
(note: $E_B < 0$).
The scattering phase shift $\delta_{l=0}(E)$ is related to the scattering
length $b$ and the width of the resonance by the following 
well--known equations:
\begin{equation}
k \cdot b = - \sin{[\delta_{l=0}(E=25~{\rm{meV}})]}
\label{eq:phaselength}
\end{equation}
and
\begin{equation}
\tan{[\delta_{l=0}(E)]} = \frac{\Gamma(E)}{2(E_B-E)}
\label{eq:phasewidth}
\end{equation}
where $k$ is the wave number of the s--wave at $E=25~{\rm{meV}}$.

Third, the experimental width $\Gamma^{\rm{exp}}$ is calculated from
Eqs.~\ref{eq:phaselength} and \ref{eq:phasewidth} 
using the experimentally determined
scattering length $b^{\rm{exp}}$ \cite{sears92,koester91}.
The SF which is a measure of the single--particle strength
is calculated from Eq.~\ref{eq:sfwidth} by the ratio of $\Gamma^{\rm{exp}}$
and $\Gamma^{\rm calc}_{\rm sp}$ at the thermal energy
$E = 25~{\rm{meV}}$.

Our new results are listed in Tab.~\ref{tab:tab1}. The main uncertainties in
this procedure are given by the experimental uncertainties of the
experimental scattering lengths. The uncertainties from different
potential parameterizations are practically negligible. The results agree
well with different transfer experiments.

The theoretical SF were calculated from the shell model
with the code OXBASH \cite{bro84}.
Since we need the spectroscopic factors for a 2s$_{1/2}$ transition,
one--particle one--hole excitations have to be taken into account for
the C--isotopes. We used the interaction WBN of Warburton and Brown
\cite{war92} for this purpose. 
For the $^{16}$N states we took the interaction ZBMI \cite{zuker68};
the results for $^{16}$N were already published in Ref.~\cite{meiss96}.
The spectroscopic factors for the
O-- and Ne--isotopes were calculated with the USD interaction of
Wildenthal \cite{wil84}.
The shell model SF agree well with the experimental 
SF derived from scattering lengths.

In the case of $^{14}$C = $^{13}$C $\otimes$ $\rm{n}$ the SF for two bound
$s$--wave states ($J^{\pi} = 0^-, 1^-$) 
can be determined, because this procedure can be applied
to both channel spins $S = 0$ and $S = 1$. The relevant scattering lengths
can be derived from the coherent and the incoherent scattering length on 
$^{13}$C. The same arguments hold for the case 
$^{16}$N = $^{15}$N $\otimes$ $\rm{n}$. However, for the nucleus $^{16}$N
the agreement between the experimental SF derived from our method and from 
a (d,p) transfer experiment is quite poor whereas the theoretical SF agree
well with our new SF.

In the cases of $^{37}$Ar = $^{36}$Ar $\otimes$ $\rm{n}$ 
and $^{41}$Ar = $^{40}$Ar $\otimes$ $\rm{n}$ subthreshold resonances
at $E = - 10~{\rm{keV}}$ ($E_x = 8778~{\rm{keV}}$)
and $E = - 1~{\rm{keV}}$ ($E_x = 6098~{\rm{keV}}$) \cite{mu81}
determine the scattering lengths. Unfortunately, the relatively
small SF of these states were not determined 
experimentally \cite{sen74,sen75}; a calculation
of these SF is very difficult because the neutron is located in the 3$s_{1/2}$
shell.

Finally, for the system $^{15}$C = $^{14}$C $\otimes$ $\rm{n}$ we can invert
the procedure to predict the experimentally unknown
scattering length of $^{14}$C from the SF of the
$^{15}$C groundstate ($1/2^+$). The SF is well--known both from
transfer experiments \cite{murillo94,cecil75} and from the shell model:
we adopt $C^2 S = 1.0 \pm 0.05$. The resulting scattering length is
$b = 7.257 \pm 0.369~\rm{fm}$. An experimental verification of this prediction
is desirable.

In conclusion, this method for the calculation of SF works well
for several light and intermediate nuclei. Because of
the model independence the SF presented in this work can be 
used as a benchmark for SF derived from transfer reactions or determined
by shell model calculations.

\acknowledgments
We would like to thank Drs.~H.~Beer, G.~Staudt, and V.~K\"olle for the
stimulating discussions during the preparation of the paper.
This work was supported by
Fonds zur F\"orderung der wissenschaftlichen Forschung 
(FWF project S7307--AST) and
Deutsche Forschungsgemeinschaft (DFG project Mo739).

\begin{table}
\caption{
	Spectroscopic factors of bound s--wave states of
	$^{13}$C, $^{14}$C, $^{16}$N,
	$^{17}$O, $^{19}$O, $^{23}$Ne, $^{37}$Ar, and $^{41}$Ar
	derived from the scattering length, from different
	transfer experiments,
	and from the shell model.
}

{
\begin{tabular}{|c|c|c|c|c|cc|cc|}
nucleus	& $J^{\pi}$	& $E_x~{\rm{(keV)}}$	& $q = 2n + l$
	& $C^2 S$ \tablenotemark[1]
	& $C^2 S^{\rm{exp}}$	& Ref.
	& $C^2 S^{\rm{calc}}$	& Ref. \\
\hline
$^{13}$C	& $1/2^+$	& 3089		& 2
		& 0.966 $\pm$ 0.015
		& 0.65 -- 1.2	& \protect\cite{ohnuma85,darden73,cook86}
		& 0.85		& \tablenotemark[1] \\
$^{14}$C	& $1^-$		& 6094		& 2
		& 0.894 $\pm$ 0.020
		& 0.43 -- 0.87	& \protect\cite{peter84,datta78,cook87}
		& 0.76 -- 0.85	
	& \tablenotemark[1], \protect\cite{lie72,jaeger71,mill75} \\
$^{14}$C	& $0^-$		& 6903		& 2
		& 0.931 $\pm$ 0.020
		& 1.02		& \protect\cite{peter84}
		& 0.96 -- 1.00	
	& \tablenotemark[1], \protect\cite{lie72,jaeger71,mill75} \\
$^{16}$N	& $0^-$		& 120		& 2
		& 1.012 $\pm$ 0.020
		& $\approx 0.46$& \protect\cite{bohne72}
		& 0.95		& \protect\cite{meiss96} \\
$^{16}$N	& $1^-$		& 397		& 2
		& 0.969 $\pm$ 0.020
		& $\approx 0.52$& \protect\cite{bohne72}
		& 0.96		& \protect\cite{meiss96} \\
$^{17}$O	& $1/2^+$	& 870		& 2
		& 0.989 $\pm$ 0.010
		& 0.45 -- 1.96	
	& \protect\cite{kocher71,cava72,cooper73,west75,rae78,moto75,yasue92}
		& 1.0	
	& \tablenotemark[1], \protect\cite{yasue92} \\
$^{19}$O	& $1/2^+$	& 1472		& 2
		& 0.919 $\pm$ 0.020
		& 0.86 -- $\approx\,$1	& \protect\cite{yasue92,sen73}
		& 0.7 -- 0.9	
	& \tablenotemark[1], \protect\cite{yasue92,mcgrory73} \\
$^{23}$Ne	& $1/2^+$	& 1017		& 2
		& 0.698 $\pm$ 0.030		
		& 0.37 -- 0.70	& \protect\cite{lutz67,nann69,howard70}
		& 0.654		& \tablenotemark[1] \\
$^{37}$Ar	& $1/2^+$	& 8789		& 4
		& 0.530 $\pm$ 0.010
		& --		& 
		& --		& \\
$^{41}$Ar	& $1/2^+$	& 6098		& 4
		& 0.180 $\pm$ 0.010
		& -- 		& 
		& --		& \\
\end{tabular}
}
\tablenotetext[1]{this work}
\label{tab:tab1}
\end{table}

\end{document}